\documentclass[review]{elsarticle}
\usepackage{float}
\usepackage{caption}
\usepackage{amsmath}
\usepackage{bm}
\usepackage{stmaryrd}
\usepackage{amssymb}
\usepackage{graphicx}
\usepackage{textcomp}
\usepackage{calrsfs}
\usepackage{yfonts}
\usepackage{color}
\usepackage{tikz}
\usepackage{lineno,hyperref}
\usepackage{amsmath}
\modulolinenumbers[5]
\newcommand{\tr}{\rm Tr}
\newcommand{\bea}{\begin{eqnarray}}
\newcommand{\eea}{\end{eqnarray}}
\newcommand{\be}{\begin{equation}}
\newcommand{\ee}{\end{equation}}
\newcommand{\nn}{\nonumber}
\newcommand{\la}{\langle}
\newcommand{\ra}{\rangle}
\newcommand{\akd}{\hat{a}_{\mathbf{k}}^\dag}
\newcommand{\ak}{\hat{a}_{\mathbf{k}}}

\newcommand{\kb}{\mathbf{k}}
\newcommand{\ok}{\omega_{\mathbf{k}}}
\newcommand{\xik}{\xi_\mathbf{k}}
\newcommand{\alfak}{\alpha_\mathbf{k}}
\newcommand{\alfask}{\alpha^*_\mathbf{k}}
\newcommand{\etak}{\eta_\mathbf{k}}

\newcommand{\xisk}{\xi^*_\mathbf{k}}
\newcommand{\Dk}{\hat{D}_{\mathbf{k}}}
\newcommand{\Dkd}{\hat{D}^\dag_{\mathbf{k}}}
\newcommand{\hh}{\hat{H}}
\newcommand{\Uk}{\hat{U}_\mathbf{k}}
\newcommand{\Ukd}{\hat{U}^\dag_\mathbf{k}}

\newcommand{\intt}{\int\limits_{0}^t}

\newcommand{\akdot}{\dot{\hat{a}}_{\mathbf{k}}}
\newcommand{\Gnk}{G_{\mathbf{k}} (\nu,t)}
\newcommand{\Gnkn}{G^{(n)}_{\mathbf{k}} (\nu,t)}
\newcommand{\Gnkth}{G^{th}_{\mathbf{k}} (\nu,t)}
\newcommand{\Gnkc}{G^{coh}_{\mathbf{k}} (\nu,t)}
\newcommand{\nnum}{\nonumber}
\newcommand{\rok}{\hat{\rho}_\mathbf{k}}
\newcommand{\rokt}{\tilde{\rho}_\mathbf{k}}
\newcommand{\F}{\mathcal{F}}
\newcommand{\zk}{\zeta_{\mathbf{k}}}

\newcommand{\half}{\frac{1}{2}}
\begin{document}

\begin{frontmatter}

\title{Many-body work distributions}

\author{Fardin Kheirandish\fnref{f.kheirandish@uok.ac.ir}}
\address{Department of Physics, Faculty of Science, University of Kurdistan, P.O.Box 66177-15175, Sanandaj, Iran}
\begin{abstract}
The work distribution function for a non-relativistic, non-interacting quantum many-body system interacting with classical external sources is investigated. Exact expressions for the characteristic function corresponding to the work distribution function is obtained for arbitrary switching function and coupling functions. The many-body frequencies are assumed to be generally time-dependent in order to take into account the possibility of moving the boundaries of the system in a predefined process linking the characteristic function to the fluctuation-induced energies in confined geometries. Some limiting cases are considered and discussed.
\end{abstract}
\begin{keyword}
Many-body system\sep Work distribution function \sep Characteristic function \sep Switching function \sep Coupling functions \sep External source
\end{keyword}
\end{frontmatter}
%
\section{Introduction}
While discussing quantum thermodynamics of a small system, quantum fluctuations being of the same order of magnitude as expectation values, play an important role. Fluctuation theorems describe the connection between non-equilibrium fluctuations and thermal equilibrium states of small systems. They also describe the nonlinear response of a
system under the influence of a time-dependent external source \cite{Jar2,Jar4}. Various fluctuation and work theorems have been studied \cite{Bochkov,Evans}. Fluctuation relations were initially derived for classical systems \cite{Jar2,Jar1,Jar3,Crooks} and experimentally confirmed in some classical systems \cite{Collin,Ex1,Ex2,Ex3}. The generalization of the fluctuation theorems to the realm of small quantum mechanical systems has led to considerable progress in formulating quantum thermodynamics \cite{Q1,Q2,Q3,Q4,Q5,Q6,Q7,Campisi}.

An important quantity in this context is the amount of work that can be extracted or done on a quantum system in a nonequilibrium process. But contrary to classical statistical mechanics work is a subtle concept in quantum thermodynamics and it is not an observable \cite{Talkner}. Therefore, indirect methods should be applied to measure work. The most established method is the two-point measurement scheme which is based on projective measurements of energy at the end-points of a process \cite{Esposito}.

Projective measurements when applied to relativistic quantum systems cause inconsistencies. Since due to the relativistic structure, the projective measurements can not be localized \cite{Redhead} and lead to the possibility of superluminal signaling \cite{Sorkin}, so they should be banished from any consistent relativistic quantum theory \cite{Sorkin,Dowker,Ben}. An effective way to remove these inconsistencies is by coupling the main system locally to auxiliary systems or ancillae like detectors, atoms or qubits. Using a controllable ancilla interacting with the main system have enabled the first experimental characterization of quantum fluctuation relations \cite{Souza}. In a recent work \cite{Ortega}, the work distributions on a relativistic quantum field has been investigated in the framework of Ramsey interferometric scheme \cite{Souza,Dorner,Mazzola}.

The Hamiltonian Eq. (\ref{1}) is a collection of non-interacting quantum driven harmonic oscillators specified by an index $\kb$ that can be a collection of particle characteristics like polarization ($\lambda $), spin ($s$) or wavenumber ($\vec{k}$),  ($\kb=\{\lambda,s,\vec{k}\}$). The distribution of work for a driven quantum harmonic oscillator with a time-independent frequency has been investigated in \cite{Q7,Os1,Os2,Talkner2008}.

In this letter, we investigate the work distribution function for a non-relativistic quantum many-body system interacting with classical external sources. The duration of the interaction of external sources with the many-body system is controlled by inserting a switching function into the interaction part of the Hamiltonian. For generality, the many-body frequencies are assumed to be time-dependent. This can arise as a result of applying external sources to the system known as Lamb-shifts or due to moving the boundaries of the system. Here we ignore from the Lamb-shifts caused by external sources and just consider the effects due to moving boundaries. In the absence of external sources, the change in the energy of the system due to the variations in the boundary of the system is studied in the context of the Casimir physics \cite{Bordag} and we have shown that it is connected to the characteristic function. We have found an exact characteristic function for arbitrary coupling functions and switching function and considered two interesting limiting cases.
\section{The model}
Consider a non-relativistic, non-interacting quantum many-body system interacting with external sources described by the second quantized Hamiltonian
\bea\label{1}
\hh &=& \sum_{\kb} \underbrace{\hbar\ok(t)\,(\akd \ak+\half)}_{\hat{H}^{\kb}_0 (t)} + \sum_{\kb} \underbrace{\hbar G(t)(\F^*(\kb) \ak+\F(\kb)\akd)}_{\hat{V}^{\kb} (t)},\nn\\
&=& \sum_{\kb} \hat{H}^{\kb} (t),\,\,\,\,\,(\hat{H}^{\kb} (t)=\hat{H}^{\kb}_0 (t)+\hat{V}^{\kb} (t)),
\eea
where $\F(\kb)(\F^* (\kb))$ are coupling functions between many-body system and external sources and $G(t)$ is a switching time-dependent function defined by
\be\label{2}
G(t)=\left\{
       \begin{array}{ll}
         g(t), & 0\leq t\leq \tau; \\
         0, & t>\tau.
       \end{array}
     \right.
\ee
Note that the Hamiltonian Eq. (\ref{1}) is the sum of independent Hamiltonians $\hat{H}^{\kb} (t)$, ($[\hat{H}^{\kb} (t),\hat{H}^{\kb'} (t)]=0$). The possibility of moving the boundaries of the many-body system and its effect on the energy spectrum of the system has been taken into account by considering the frequencies $\ok (t)$ to be generally time-dependent. So, one can affect the many-body system by two processes: (i) By applying external classical sources, like the interaction of external electromagnetic fields with the system. (ii) By moving the boundaries of the system, for example in the case of a bosonic gas confined between parallel conducting plates, the distance between the plates can be considered as a varying parameter.

Here we have adopted the lab quantization, so the wave number in $\kb$, denoted by $\vec{k}$ takes on discrete values, if the volume of the lab tends to infinity, the discrete sums over $\vec{k}$ should be replaced by the integral
\be\label{3}
\sum_{\vec{k}}\rightarrow \int \frac{d^3 \vec{k}}{(2\pi)^3}.
\ee
\section{The time-evolution operator}
Let us focus on the $\kb$th subsystem in the Hamiltonian Eq. (\ref{1}), from the Heisenberg equations of motion, we find for the annihilation operator $\ak$
\be\label{4}
\akdot=-i\ok (t) \ak-i G(t)\, \F (\kb),
\ee
with the formal solution
\be\label{5}
\ak (t)=e^{-i\zk (t)} \big(\ak (0)-\xik (t)\big),
\ee
where we have defined
\bea\label{6}
\zk (t) &=& \intt dt'\,\ok (t'),
\eea
and
\bea\label{7}
\xik (t) &=& i \F (\kb)\,\intt dt'\,e^{i\zk (t')}\,G(t').
\eea
From Eq. (\ref{2}) it is clear that $\xik (t)$ is time-independent for $t\geq \tau$,  ($\xik(t)=\xik(\tau)$).

An important operator in quantum optics terminology is the displacement operator
\be\label{8}
\Dk (\xik)=e^{\xik (t)\akd (0)-\xisk (t)\ak},
\ee
with displacement parameter $\xik (t)$, which acts on the annihilation operator as
\be\label{9}
\Dk (\xik)\ak (0)\Dkd (\xik)=\ak (0)-\xik (t).
\ee
The displacement operator when applied to the vacuum state $|0\ra$ produces a coherent state $|\xik (t)\ra=\Dk (\xik)|0\ra$ which is an eigenket of the annihilation operator $\ak |\xik(t)\ra=\xik (t) |\xik (t)\ra$.

In Heisenberg picture, the time-evolution of the annihilation operator is given by
\be\label{10}
\ak (t)=\Ukd (t) \ak (0) \Uk (t),
\ee
where $\Uk (t)$ is the time-evolution operator corresponding to the kth subsystem with the Hamiltonian $\hat{H}^{\kb} (t)$. One easily finds (\ref{ApA})
\be\label{11}
\Uk (t)= e^{i \theta (t)}\,e^{-i\zk (t)\,(\akd \ak+1/2)}\,\Dkd (\xik).
\ee
The time-dependent Hamiltonian $\hat{H}_0^{\kb} (t)$ satisfies $[\hat{H}_0^{\kb} (t)$, $\hat{H}_0^{\kb} (t')]=0$, therefore,
in the absence of the external sources ($\F(\kb)=0$), the time-evolution operator for the $\kb$th subsystem is given by
\bea\label{12}
\hat{U}_{0\kb} (t) &=& e^{-\frac{i}{\hbar}\int_0^t \hat{H}_0^{\kb}(t')\,dt'},\nn\\
              &=& e^{-i\zk (t)\,(\akd \ak+1/2)}.
\eea
The time-evolution operator for the $\kb$th subsystem in the interaction picture is by definition
\bea\label{13}
\hat{U}_I(t) &=& \hat{U}^\dag_{0\kb} (t) \Uk(t),\nn\\
             &=& e^{i \theta (t)}\,e^{i\zk (t)\,(\akd \ak+1/2)}e^{-i\zk (t)\,(\akd \ak+1/2)}\,\Dkd (\xik)\nn\\
             &=& e^{i \theta (t)}\,\Dkd (\xik).
\eea
that is the time-evolution operator in the interaction picture is the adjoint of the displacement operator.
\section{Characteristic function for a single subsystem}
A quantum system responds to external sources that can be characterized by finding the variation of the energy of the system including the interaction with external sources. In this context, the work $W$ is defined as the difference between the final and initial energies of the system plus interaction.
Let us focus on the $\kb$th subsystem with Hamiltonian $\hat{H}^{\kb} (t)$, and let $E_i,\,(i=1,2,\cdots)$ be the energy spectrum of the Hamiltonian $\hat{H}^{\kb} (0)$, at $t=0$, and let $E'_i,\,(i=1,2,\cdots)$, be the energy spectrum of the Hamiltonian $\hat{H}^{\kb} (t)$ at $t>0$. To find the energy spectrum of the Hamiltonian $\hat{H}^{\kb} (t)$, we consider the unitary transformation
\bea\label{14}
\Dk (\alfak)H^{\kb}_0 (t)\Dkd (\alfak) &=& e^{\alfak \akd -\alfask \ak}\hbar\ok (t)[\akd\ak+1/2]e^{\alfask \ak -\alfak \akd},\nn\\
             &=& \hbar\ok(t)\,[(\akd-\alfask)(\ak-\alfak)+\half],\nn\\
             &=& \hbar\ok(t)\,(\akd \ak+\half)-\hbar\ok(t)(\alfask \ak+\alfak \akd-|\alfak|^2),\nn\\
\eea
by setting
\bea\label{15}
\alfak (t)= -\frac{G(t)}{\ok (t)}\,\F(\kb),
\eea
we will find
\bea\label{16}
 \hat{H}^{\kb} (t) &=& \hbar\ok(t)\,(\akd \ak+\half)+\hbar G(t)(\F^*(\kb) \ak+\F(\kb)\akd),\nn\\
 &=& \Dk (\alfak)H^{\kb}_0 (t)\Dkd (\alfak)-\hbar\ok (t)|\alfak|^2,
\eea
where $\alfak$ is given by Eq. (\ref{15}). According to Eq. (\ref{16}), if $|\psi_n\ra$ is an eigenket of the Hamiltonian $H^{\kb}_0 (t)$ with eigenvalue $E_n=\hbar\ok (t) (n+1/2)$ then $ \Dk (\alfak)|\psi_n \ra$ is an eigenket of the Hamiltonian $\hat{H}^{\kb} (t)$ with eigenvalue
\be\label{17}
E'_n=\hbar\ok (t)(n+\half)-\frac{\hbar G^2 (t) |\F (\kb)|^2}{\ok (t)}.
\ee
The characteristic function corresponding to the $\kb$th subsystem is denoted by $\Gnk$ and provides a complete statistical description of the work performed on or extracted from the subsystem. The characteristic function is defined as the Fourier transform of the probability density of the work $P_k (W,t)$ \cite{Esposito}
\be\label{18}
P_k (W,t)=\sum_{i,j} p_0 (E_i)p_t (|E_i\ra\rightarrow |E'_j\ra)\,\delta(W-E'_j+E_i),
\ee
where $p_0(E_i)$ is the probability that the kth subsystem be in the eigenstate $|E_i\ra$ with the eigenvalue $E_i$ at $t=0$, and $p_t (|E_i\ra\rightarrow |E'_j\ra)$ is the transition-probability from the initial state $|E_i\ra$ to the final state $|E'_j\ra$ at time $t$. Let the diagonal part of the initial density matrix $\rok (0)$, in the basis of energy eigenkets $\{|E_i\ra\}$, be denoted by
\bea\label{19}
&& \rokt (0) = \sum_i p(E_i)|E_i\ra\la E_i|,\nn\\
&& \sum_i p(E_i) = 1,
\eea
then from the definition of the characteristic function
\bea\label{20}
\Gnk &=& \int_{-\infty}^{\infty} dW\,e^{i W \nu} P_k (W,t)=\la e^{i\nu W}\ra_{\kb},\nnum\\
&=& \sum_{i,j} p_0 (E_i)p_t (|E_i\ra\rightarrow |E'_j\ra)\,e^{i\nu (E'_j-E_i)}.
\eea
and the transition probability
\be\label{21}
p_t (|E_i\ra\rightarrow |E'_j\ra)=|\la E'_j|\hat{U}_{\kb} (t)|E_i\ra|^2,
\ee
we can rewrite Eq. (\ref{20}) as (\ref{ApB})
\bea\label{22}
\Gnk &=& \tr\big[e^{-i\nu \hat{H}_0^{\kb} (0)}\,\hat{U}_{\kb}^{\dag}(t)\,e^{i\nu \hat{H}^{\kb} (t)}\, \hat{U}_{\kb}(t)\, \rokt (0)\big].
\eea
\subsection{Example 1.}
Let the initial state be the number state $ \rok (0)=|n\ra\la n|=\rokt(0)$, then
\bea\label{23}
\Gnkn &=& \tr\bigg[e^{-i\nu \hat{H}_0^{\kb}(0)}\,\hat{U}^{\dag}_{\kb} (t)\,e^{i\nu \hat{H}^{\kb} (t)}\,\hat{U}_{\kb}(t)|n\ra\la n|\bigg],\nn\\
     &=& \la n|e^{-i\nu \hat{H}_0^{\kb}(0)}\,\hat{U}^{\dag}_{\kb} (t)\,e^{i\nu \hat{H}^{\kb} (t)}\,\hat{U}_{\kb}(t)|n\ra,
\eea
after straightforward calculations, one finds (\ref{ApC})
\bea\label{24}
\Gnkn &=& e^{-i\nu\hbar\ok (t)|\alfak|^2}e^{|\etak|^2 (e^{i\nu\hbar\ok (t)}-1)}e^{i\nu\hbar\triangle_{\kb} (t)(n+1/2)}\nn\\
&& \times\,L_n \bigg(4|\etak|^2\,\sin^2\frac{\nu\hbar\ok (t)}{2}\bigg),
\eea
where $L_n (x)$ is the Laguerre polynomial of degree $n$ and
\bea\label{25}
\triangle_{\kb} (t) &=& \ok (t)-\ok (0),\nn\\
\etak (t) &=& \xik (t) + \alfak (t)\,e^{i\zk (t)}.
\eea
In the special case where the boundaries of the system are fixed, the frequencies are time-independent ($\ok (t)=\ok (0)$), so $\triangle_{\kb} (t)=0$ and the Eq. (\ref{24}) tends to the same result reported in \cite{Talkner2008}.

Now let us find the work distribution function corresponding to the characteristic function Eq.(\ref{24}). For this purpose, we rewrite Eq. (\ref{24}) as follows
\bea\label{wdf}
\Gnkn &=& e^{-i\nu\hbar\ok (t)|\alfak|^2}e^{-|\etak|^2} e^{|\etak|^2\,e^{i\nu\hbar\ok (t)}}e^{i\nu\hbar\triangle_{\kb} (t)(n+1/2)}\nn\\
&& \times\,\sum_{l=0}^{n}\binom{n}{l}\frac{(-4|\etak|^2\,\sin^2(\nu\hbar\ok(t)/2))^l}{l!},
\eea
where we used the following expansion for the Laguerre polynomial of the order n
\be
L_n (x)=\sum_{l=0}^n \binom{n}{l}\frac{(-x)^l}{l!}.
\ee
The work distribution function can be obtained from  Eq. (\ref{wdf}) by taking the inverse Fourier transform. After straightforward calculations, we will find (\ref{ApD})
\bea\label{PkW}
P^{(n)}_k(W,t) &=& \sum_{s=-n}^\infty \sum_{l=0}^n \sum_{p=\mbox{max} (l-s,0)}^{2l} \binom{n}{l}\binom{2l}{p}(-1)^p\,\frac{e^{-|\etak|^2}|\etak|^{2(s+p)}}{l! \,(s+p-l)!}\nn\\
         && \times\,\delta\Big([s+n+1/2]\hbar\triangle_{\kb}(t)+s\hbar\ok(0)-\hbar\ok(t)|\alfak|^2-W\Big).\nn\\
\eea
From Eq. (\ref{PkW}) the weight functions are defined by \cite{Talkner2008}
\bea\label{qsn}
q_s (n,|\etak|^2)=\sum_{l=0}^n \sum_{p=\mbox{max} (l-s,0)}^{2l} \binom{n}{l}\binom{2l}{p}(-1)^p\,\frac{e^{-|\etak|^2}|\etak|^{2(s+p)}}{l! \,(s+p-l)!}.
\eea
In Eq. (\ref{qsn}) the dimensionless quantity $|\etak|$ corresponds to the rapidity parameter defined in \cite{Talkner2008} for $\triangle_{\kb}(t)=0$. Note that for $t>\tau$  we have $G(t)=\alfak=0$ (see Eqs. (\ref{2}), (\ref{15})), therefore, from Eqs. (\ref{25}), (\ref{7}) we have
\bea\label{rapi}
z &=& |\etak|=|\xik (t) + \alfak (t)\,e^{i\zk (t)}|,\nn\\
&=& |\xik(\tau)|^2=\Big|\F (\kb)|^2\,|\int_0^{\tau} dt'\,e^{i\zk (t')}\,G(t')\Big|^2.
\eea
For the case where the frequencies are time-independent $\ok(t)=\ok(0)$, Eq. (\ref{rapi}) can be simplified as
\be\label{srapi}
z=|\F (\kb)|^2\,|\tilde{G}(\ok(0))|^2,
\ee
where $\tilde{G}(\ok(0))$ is the Fourier transform of the switching function $G(t)$.

The weight functions $q_s(3,z)$ in terms of the rapidity $z$ have been depicted in Fig.1 for $s=-3,...,3$, ($\rho_k (0)=|3\ra\la 3|$) and in Fig.2 for $s=0,1,2,3$, ($ \rho_k (0)=|0\ra\la 0|$). In both figures the location of the maximum of weight function has been shifted to the larger values of $z$ by increasing $|s|$.
\begin{figure}[htb!]
\centering
  \includegraphics[scale=0.30]{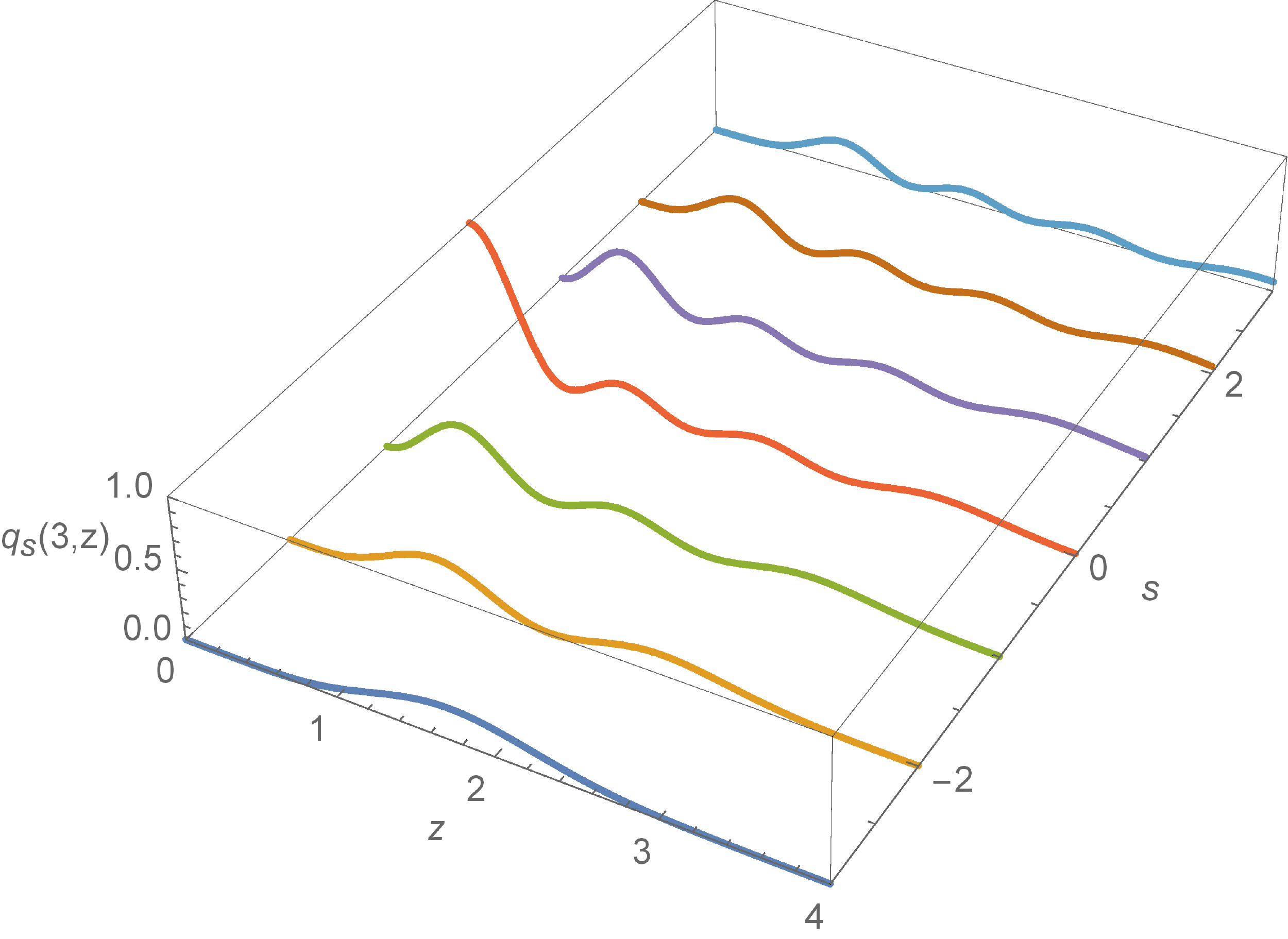}
  \caption{(Color online) The weight functions $q_s(3,z)$ in terms of the rapidity $z$ for the initial state with $n=3$ and $s=-3,...,3$.}
\end{figure}
\begin{figure}[htb!]
\centering
  \includegraphics[scale=0.30]{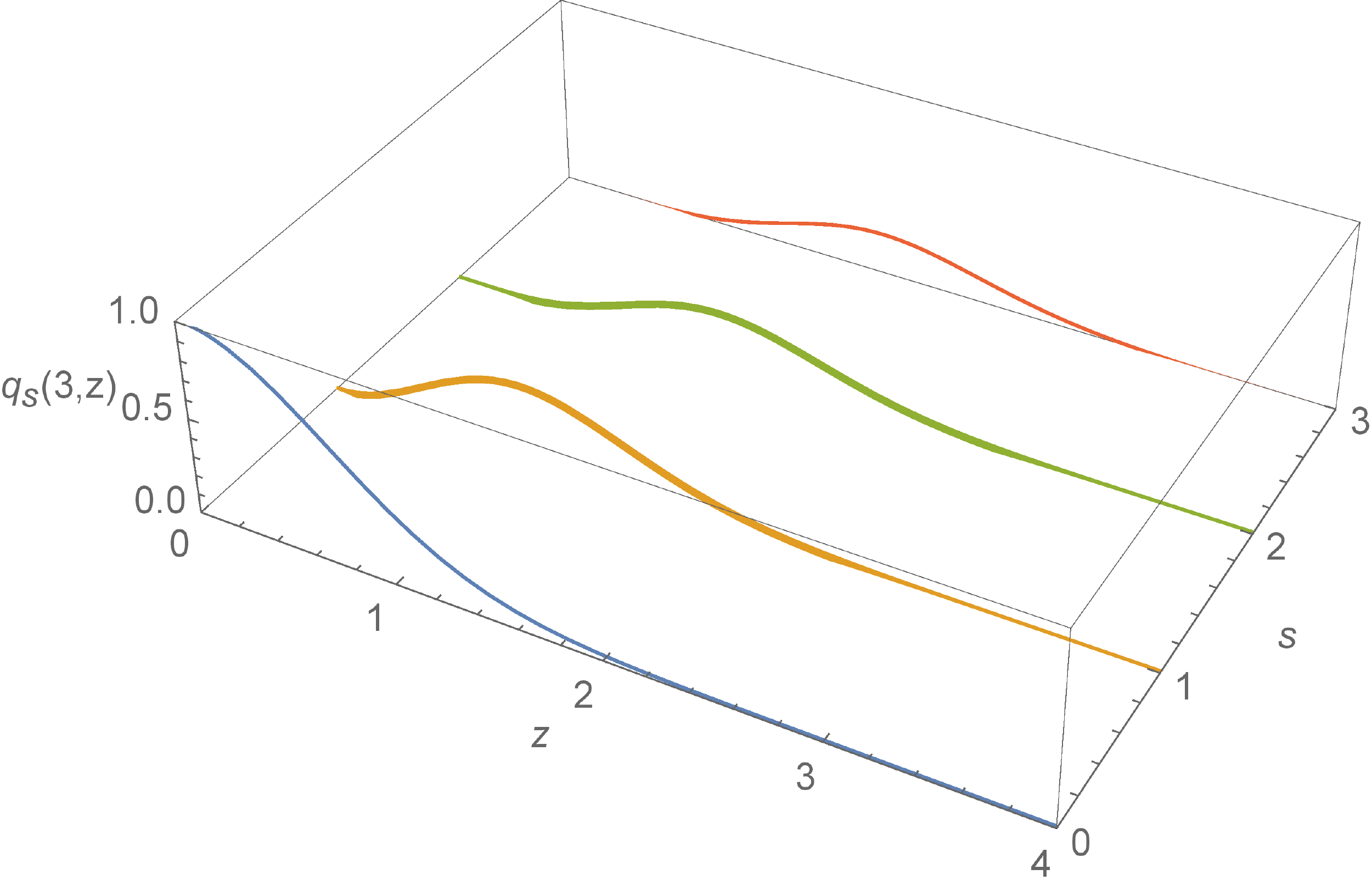}
  \caption{(Color online) The weight functions $q_s(0,z)$ in terms of the rapidity $z$ for the initial state with $n=0$ and $s=0,1,2,3$.}
\end{figure}
\subsection{Example 2.}
Let the initial density matrix be a thermal state with inverse temperature $\beta=1/k_B T$ ($k_B$ is the Boltzmann constant)
\bea\label{26}
\rok (0) &=& \rokt(0)= \sum_i p(E_i)|E_i\ra\la E_i|,\,\,\, p(E_i)=\frac{e^{-\beta E_i}}{z_k (0)},\nn\\
&=& \frac{e^{-\beta \hat{H}^k_0 (0)}}{z_k (0)},
\eea
where $z_k (0)$ is the partition function
\bea\label{27}
z_k (0) &=& \tr[e^{-\beta\hat{H}^{\kb}_{0}(0)}],\nn\\
&=& \frac{e^{-\frac{\beta\hbar\ok (0)}{2}}}{1-e^{-\beta\hbar\ok (0)}},
\eea
then we will find the characteristic function $\Gnkth$ as (\ref{ApD})
\bea\label{28}
\Gnkth &=& e^{-i\nu\hbar\ok (t) |\alfak|^2}\,e^{|\etak (t)|^2(e^{i\nu\hbar\ok (t)}-1)}e^{\frac{i\nu\hbar\triangle_{\kb} (t)}{2}}\nn\\
&& \times \frac{(1-e^{-\beta\hbar\ok (0)})\,e^{-\frac{4|\etak (t)|^2\sin^2 (\frac{\nu\hbar\ok (t)}{2})}{e^{\beta\hbar\ok (0)}e^{-i\hbar\nu\triangle_{\kb} (t)}-1}}}{1-e^{i\hbar\nu\triangle_{\kb} (t)}\,e^{-\beta\hbar\ok (0)}},
\eea
which is in agreement with the result reported in \cite{Talkner2008} when $\triangle_{\kb}=0$.

An interesting case arises when the external sources are switched off ($\xik=0$) but the boundaries of the many-body system change with time in a predefined process. In this case $\alfak=\etak=\xik=0$, therefore,
\bea\label{29}
G_{\kb}^{bc} (\nu,t) &=& \frac{e^{i\hbar\nu\triangle_{\kb} (t)/2}\,(1-e^{-\beta\hbar\ok (0)})}{1-e^{i\hbar\nu\triangle_{\kb} (t)}e^{-\beta\hbar\ok (0)}}.
\eea
To find the work distribution function we rewrite Eq. (\ref{29}) as
\bea\label{wdfbc}
G_{\kb}^{bc} (\nu,t) &=& (1-e^{-\beta\hbar\ok(0)})\,e^{i\hbar\nu\frac{\triangle_{\kb} (t)}{2}}\,\sum_{r=0}^\infty e^{ir\hbar\nu\triangle_{\kb} (t)}\,e^{-\beta r\hbar\ok(0)},
\eea
and by taking the inverse Fourier transform we find
\bea
P_{\kb}^{bc} (W,t) &=& \sum_{r=0}^{\infty} (1-e^{-\beta\hbar\ok(0)})\,e^{-\beta r\hbar\ok(0)}\,\delta\Big[\big(r+\frac{1}{2}\big)\hbar\triangle_{\kb} (t)-W\Big],
\eea
with the weight functions
\be
q_r=(1-e^{-\beta\hbar\ok(0)})\,e^{-\beta r\hbar\ok(0)}.
\ee
The probabilities $q_r$ correspond to the initial thermal state Eq. (\ref{26}) and the sign of the work $W$ depend on the sign of $\triangle_{\kb} (t)$. If work is done on a system the eigenfrequency increases ($\triangle_{\kb} (t)>0$) and when work is delivered by the system the eigenfrequency decreases ($\triangle_{\kb} (t)<0$).
\subsection{Example 3.}
Let the initial state be a coherent state $\rok (0)=|\alpha\ra\la \alpha|$,
\be\label{cohe}
|\alpha\ra=e^{-\frac{|\alpha|^2}{2}}\sum_{n=0}^\infty \frac{\alpha^n}{\sqrt{n!}}|n\ra,
\ee
then
\bea
\rokt(0)=e^{-|\alpha|^2}\sum_{n=0}^\infty \frac{|\alpha|^{2n}}{n!}|n\ra\la n|,
\eea
and the characteristic function is
\bea\label{gnkc}
\Gnkc=\sum_{n=0}^\infty e^{-|\alpha|^2} \frac{|\alpha|^{2n}}{n!}\,\Gnkn,
\eea
where $\Gnkn$ is given by Eq. (\ref{wdf}). Therefore, the corresponding work distribution function is
\bea\label{wdfc}
P^{coh}_{\kb} (W,t)=\sum_{n=0}^\infty e^{-|\alpha|^2} \frac{|\alpha|^{2n}}{n!}\,P^{(n)}_k(W,t),
\eea
where $P^{(n)}_k(W,t)$ is given by Eq. (\ref{PkW}). From Eq. (\ref{wdfc}) it is clear that the work distribution function in this case is the average of $P^{(n)}_{\kb} (W,t)$ with respect to the Poisson distribution function. By inserting Eq. (\ref{24}) into Eq. (\ref{gnkc}) we have
\bea\label{coherent}
\Gnkc &=& e^{-i\nu\hbar\ok (t)|\alfak|^2}e^{|\etak|^2 (e^{i\nu\hbar\ok (t)}-1)}\,e^{i\nu\hbar\triangle_{\kb} (t)/2}\,e^{-|\alpha|^2}\nn\\
      &\times &\, \sum_{n=0}^\infty \frac{\Big(|\alpha|^2\,e^{i\nu\hbar\triangle_{\kb} (t)}\Big)^n}{n!}\,L_n \bigg(4|\etak|^2\,\sin^2\frac{\nu\hbar\ok (t)}{2}\bigg),
\eea
now by making use of the identity \cite{Gradstein}
\be\label{identity}
J_0 (2\sqrt{xy})\,e^{y}=\sum_{n=0}^\infty \frac{y^n}{n!}\,L_n (x),
\ee
we finally find
\bea
\Gnkc &=& e^{-i\nu\hbar\ok (t)|\alfak|^2}e^{|\etak|^2 (e^{i\nu\hbar\ok (t)}-1)}\,e^{i\nu\hbar\triangle_{\kb} (t)/2}\,e^{-|\alpha|^2}\nn\\
      &\times &\,J_0 \Big(4\Big|\alpha\,\etak\,\sin\frac{\nu\hbar\ok (t)}{2}\Big|\,e^{i\nu\hbar\triangle_{\kb} (t)/2}\Big)\,e^{|\alpha|^2\,e^{i\nu\hbar\triangle_{\kb} (t)}}.
\eea
where $J_0 (x)$ is the Bessel function of order zero. The Eq. (\ref{coherent}) is in agreement with the result reported in \cite{Talkner2008} when $\triangle_{\kb}=0$. In the absence of external sources we have
\bea\label{easy}
\Gnkc = e^{i\nu\hbar\triangle_{\kb} (t)/2}\,e^{|\alpha|^2\,(e^{i\hbar\nu\triangle_{\kb} (t)}-1)}.
\eea
By taking the inverse Fourier transform of Eq. (\ref{easy}), we find the work distribution function as
\be
P^{coh}_{\kb} (W,t)=\sum_{n=0}^\infty \frac{e^{-|\alpha|^2}\,|\alpha|^{2n}}{n!}\,\delta\Big((n+\frac{1}{2})\hbar\triangle_{\kb} (t)-W\Big),
\ee
where the weight functions have a poisson distribution
\be
q_n = \frac{e^{-|\alpha|^2}\,|\alpha|^{2n}}{n!}.
\ee

\section{The characteristic function of the total system}
To find the characteristic function of the total system we can easily proof that
if the total Hamiltonian $\hat{H}$ is the sum of noninteracting Hamiltonians $\hat{H}_{\kb}$
\be\label{30}
\hat{H}=\sum_{\kb} H_{\kb},
\ee
then the characteristic function of the total system is the multiplication of the characteristic functions of the subsystems given by (\ref{ApE})
\be\label{31}
G(\nu,t)=\prod_{\kb} \Gnk.
\ee
The Eq. (\ref{31}), can be rewritten  as
\bea\label{32}
G(\nu,t)=e^{\sum\limits_{\kb}\ln\Gnk},
\eea
where $\Gnk$ is given by Eq. (\ref{22}).

Now from the Eqs. (\ref{22}), (\ref{31}), and the fact that the subsystems defined by Hamiltonians $\hat{H}^{\kb} (t)$ are non-interacting, we can generalise the Eq. (\ref{22}) as
\bea\label{33}
G(\nu,t)=\tr\,[e^{-i\nu \hat{H}_0 (0)}\,\hat{U}^{\dag} (t)\,e^{i\nu \hat{H}(t)}\, \hat{U} (t)\, \hat{\rho} (0)],
\eea
where
\bea\label{34}
\hat{\rho}(0) &=& \frac{e^{-\beta \hat{H}_0 (0)}}{\tr \,e^{-\beta \hat{H}_0 (0)}}.
\eea
Also from Eq. (\ref{20}) and its inverse, we find
\bea\label{35}
G(\nu,t) &=& \int_{-\infty}^{\infty} dW\,e^{i W \nu} P (W,t)=\la e^{i\nu W}\ra,
\eea
and
\bea\label{36}
P(W,t) &=& \int_{-\infty}^{\infty}\frac{ d\nu}{2\pi}\,e^{-i\nu W} G (\nu,t).
\eea
\section{Expectation values and cumulant function}
From the definition of the characteristic function Eq. (\ref{20}), the nth moment of work can be obtained as
\bea\label{37}
\la W^n\ra=\frac{1}{i^n}\,\frac{d^n}{d\nu^n}G(\nu,t)\big{|}_{\nu=0}.
\eea
To facilitate the calculations, let us consider the characteristic function Eq. (\ref{28}) in zero temperature ($\beta\rightarrow\infty$), in this limit, we have
\bea\label{38}
G(\nu,t)=e^{\sum\limits_k \big[|\etak (t)|^2(e^{i\nu\hbar\ok (t)}-1)+i\nu\hbar(\triangle_{\kb} (t)/2-\ok (t)|\alfak|^2)\big]}.
\eea
The first and the second moments of the work are respectively given by
\bea\label{39}
\la W\ra &=& \frac{1}{i}\,\frac{d}{d\nu}G(\nu,t)\big{|}_{\nu=0},\nn\\
&=& \sum_k \Big[|\etak (t)|^2\hbar\ok (t)+\hbar\Big(\triangle_{\kb} (t)/2-\ok (t) |\alfak|^2\Big)\Big],\nn\\
\la W^2\ra &=& \frac{1}{i^2}\,\frac{d^2}{d\nu^2}G(\nu,t)\big{|}_{\nu=0},\nn\\
&=& \sum_k |\etak (t)|^2\hbar^2\ok^2 (t)+\la W\ra^2,
\eea
with the variance
\bea\label{40}
(\Delta W)^2 &=& \la W^2\ra-\la W\ra^2,\nn\\
&=& \sum\limits_{\kb} |\etak (t)|^2\hbar^2\ok^2 (t).
\eea
In general, by inserting Eq. (\ref{32}) into the cumulant function defined by \cite{Esposito}
\bea\label{41}
C(\nu,t) &=& \ln G(\nu,t),
\eea
we find
\bea\label{42}
C(\nu,t) &=&  \sum\limits_{\kb} \ln\Gnk.
\eea
From the cumulant function, we find the nth cumulant as
\bea\label{43}
C_n= \frac{1}{i^n}\frac{\partial^n}{\partial\nu^n}\,C(\nu,t)\Big{|}_{\nu=0},
\eea
where $C_1=\la W\ra$ and $C_2=(\Delta W)^2$. If the characteristic function is given by Eq. (\ref{38}), the skewness, which determines the deviation of $P(W)$ from a Gaussian distribution, is
\bea\label{44}
\la (W-\la W\ra)^3 \ra &=& \frac{1}{i^3}\frac{\partial^3}{\partial\nu^3}\,C(\nu,n)\Big{|}_{\nu=0},\nn\\
                       &=& \sum\limits_{\kb} |\etak (t)|^2\,(\hbar\ok (t))^3.
\eea
\subsection{Casimir energy}
Consider an arbitrary fluctuating field confined in a region by imposing Dirichlet boundary conditions. The region can be contracted or expanded by varying external parameters. Casimir energy is defined as the difference between zero-point energies of the system in the presence and absence of boundaries. As an illustration, consider a photonic gas confined between parallel, neutral, perfect conductors with unit area. The external parameter in this case is the distance between conductors ($d$). The Hamiltonian of the photonic gas is like Eq. (\ref{1}), without external sources, where the discrete index $\kb$ refers to the wave number and polarization of photons. In zero temperature ($\beta\rightarrow\infty$), from Eq. (\ref{29}) we have
\be\label{45}
G_{\kb}^{bc} (\nu,t) =e^{i\nu\hbar\,\triangle_{\kb} (t)/2},
\ee
the total characteristic function is
\be\label{46}
G^{bc} (\nu,t) =e^{\sum\limits_k i\nu\hbar\,\triangle_{\kb} (t)/2},
\ee
and the work distribution function is obtained as
\bea\label{47}
P(W,t) &=& \int_{-\infty}^{\infty} \frac{d\nu}{2\pi}\,e^{-i\nu W+\sum\limits_k i\nu\hbar\,\triangle_{\kb} (t)/2},\nn\\
       &=& \delta \big(\sum\limits_k \hbar\,\triangle_{\kb} (t)/2-W\big),
\eea
with zero variance, that is
\be\label{48}
W=\sum\limits_k \hbar\,\ok (t)/2-\sum\limits_k \hbar\,\ok (0)/2,
\ee
as expected. The sum over $\kb$ in Eq. (\ref{48}) can be achieved in different ways and an explicit dependence on $d$ is obtained as \cite{Bordag}
\be\label{49}
W=-\frac{\hbar c\pi^2}{720 d^3}.
\ee

In general, by setting $\beta=i\nu$ in Eq. (\ref{33}), the quantum version of Jarzynski equality is recovered \cite{Talkner}
\bea\label{50}
G(i\beta,\tau) &=& \la e^{-\beta W}\ra,\nn\\
&=& \frac{\tr e^{-\beta \hat{H}_0 (\tau)}}{\tr e^{-\beta \hat{H}_0 (0)}}=\frac{Z(\tau)}{Z(0)},\nn\\
&=&  e^{-\beta \triangle F},
\eea
where $\triangle F=F(\tau)-F(0)$, represents the difference of free energies at times $t=\tau$ and $t=0$, or equivalently for different configurations of boundaries. Therefore,
\bea\label{51}
\triangle F=-\frac{1}{\beta}\ln[G(i\beta,\tau)].
\eea
By inserting Eq. (\ref{29} into Eq. (\ref{51}), one can find the temperature correction to the Casimir energy.
\section{Summary and conclusions}
We have considered a non-relativistic non-interacting quantum many-body system under the influence of external classical sources. The external sources are coupled to the many-body system by the coupling functions $\F(\kb)\,(\F^* (\kb))$ in a finite interval of time $\tau$ controlled by the switching function $G(t)$. The switching function and the coupling functions are arbitrary functions and one can assign them desired explicit forms depending on the characteristics of the many-body system and the external sources. We have also taken into account the possibility of moving boundaries of the many-body system in a predefined process by considering the frequencies of the system to be generally time-dependent. We have obtained the closed form expression Eq. (\ref{32}) for the characteristic function $G(\nu,t)$ corresponding to the work distribution function $P(W)$. Two interesting limiting cases are considered (i) There are external sources but the boundary of the system do not change, in this case the system frequencies are time-independent and the characteristic function of the system can be obtained using Eqs.(\ref{28}) and Eq. (\ref{32}). (ii) The external sources are switched off but the boundaries of the system can be moved in a predefined process. The characteristic function in this case is given by Eqs. (\ref{29}) and (\ref{32}). In a process where the configuration of the boundaries of the many-body system at the end points are different but initial and final states of the system are thermal equilibrium states, the change in free energy is given by Eq. (\ref{51}) which is a useful formula in the context of the fluctuation-induced forces.
\appendix
\section{Proof of Eq. (\ref{11})}\label{ApA}
By setting
\bea\label{A1}
\hat{A} &=& \frac{i\zk (t)}{2}\,(\akd \ak+\ak\akd),\nn\\
\hat{B} &=& \ak (0),
\eea
in the Baker–Campbell–Hausdorff (BCH) formula
\bea\label{A2}
e^{\hat{A}} \hat{B} e^{-\hat{A}} &=& \hat{B}+\frac{1}{1!}[\hat{A},\hat{B}]+\frac{1}{2!}[\hat{A},[\hat{A},\hat{B}]]\nn\\
&& +\frac{1}{3!}[\hat{A},[\hat{A},[\hat{A},\hat{B}]]]+\cdots,
\eea
we find
\bea\label{A3}
&& e^{\frac{i\zk (t)}{2}\,(\akd \ak+\ak\akd)} \ak (0) e^{\frac{-i\zk (t)}{2}\,(\akd \ak+\ak\akd)}\nn\\
&& =e^{-i\zk (t)} \ak (0),
\eea
and similarly by setting
\bea\label{A4}
\hat{A} &=& \xik \akd-\xik^* \ak,\nn\\
\hat{B} &=& \ak (0),
\eea
we find
\bea\label{A5}
\Dk (t) \ak (0) \Dkd (t) &=& e^{\xik \akd-\xik^* \ak} \ak (0) e^{\xik^* \ak-\xik \akd}\nn\\
&=& \ak (0)-\xik (t).
\eea
Now by making use of Eqs. (\ref{A2}) and (\ref{A5}) we have
\bea
&&  \Ukd (t) \ak (0) \Uk (t) =\Dk (t)\bigg(e^{\frac{i\zk (t)}{2}\,(\akd \ak+\ak\akd)}\, \ak (0) \nn\\
&& \times e^{\frac{-i\zk (t)}{2}\,(\akd \ak+\ak\akd)}\bigg)\Dkd (t)\nonumber \\
   &=& \Dk (t)\Big(e^{-i\zk (t)}\ak (0)\Big)\Dkd\nn\\
   &=& e^{-i\zk (t)}(\ak (0)-\xik (t))=\ak (t),
\eea
therefore, $\Uk (t)$ is the quantum propagator corresponding to the Hamiltonian $\hat{H}^{\kb} (t)$. By inserting $\Uk (t)$ into the Schr\"{o}dinger equation
\be
i\hbar\,\frac{d \Uk (t)}{dt}=\hat{H}^{\kb} (t) \,\Uk (t),
\ee
we find the phase factor $\theta(t)$ in Eq. (\ref{11}) as
\be\label{tetat}
\theta(t)= |\F(\kb)|^2\,\int_0^t dt'\,\int_0^{t'}dt''\,G(t')\,\sin[\zk (t')-\zk (t'')]\,G(t'').
\ee
\section{Proof of Eq. (\ref{22})}\label{ApB}
From Eqs. (\ref{20}), (\ref{21}) we have
\bea
\Gnk &=& \sum\limits_{i,j} p_0 (E_i) |\la E'_j|U_k (t)|E_i\ra|^2\,e^{i\nu (E'_j-E_i)},\nn\\
     &=& \sum\limits_{i,j} \la E_i|\,e^{-i\nu \hat{H}^{\kb}_0 (0)}\hat{U}^{\dag}_{\kb} (t)e^{i\nu \hat{H}^{\kb} (t)}|E'_j\ra\la E'_j|\hat{U}_{\kb} (t)\rokt (0)|E_i\ra,\nn\\
     &=& \sum\limits_{i} \la E_i|\,e^{-i\nu \hat{H}^{\kb}_0 (0)}\hat{U}^{\dag}_{\kb} (t)e^{i\nu \hat{H}^{\kb} (t)}\hat{U}_{\kb} (t)\rokt (0)|E_i\ra,\nn\\
     &=& \tr\big[e^{-i\nu \hat{H}_0^{\kb} (0)}\,\hat{U}_{\kb}^{\dag}(t)\,e^{i\nu \hat{H}^{\kb} (t)}\, \hat{U}_{\kb}(t)\, \rokt (0)\big].
\eea
\section{Proof of Eq. (\ref{24})}\label{ApC}
From Eq. (\ref{23}) we have
\bea
\Gnkn &=& e^{-i\nu \hbar\ok (0)(n+1/2)}\,\la n|\hat{U}^{\dag}_{\kb} (t)\,e^{i\nu \hat{H}^{\kb} (t)}\, \hat{U}_{\kb} (t)\, |n\ra,
\eea
also from Eq. (\ref{16}) we have
\bea
\hat{U}^{\dag}_{\kb} (t)\,\hat{H}^{\kb} (t)\, \hat{U}_{\kb} (t) = \underbrace{\hat{U}^{\dag}_{\kb} (t) \Dk (\alfak)}_{\hat{V}^{\dag}} \hat{H}^{\kb}_0 (t)\underbrace{\Dkd (\alfak)\hat{U}_{\kb} (t)}_{\hat{V}}-\hbar\ok (t)|\alfak|^2,
\eea
so
\bea
\Gnkn &=& e^{-i\nu\hbar\,\ok(t)|\alfak|^2}\,e^{-i\nu\hbar\,\ok (0) (n+1/2)}\nn\\
     && \times \sum\limits_{m=0}^\infty \la n|\hat{V}^{\dag}|m\ra\la m|\hat{V}|n\ra e^{i\nu\hbar\,\ok (t) (m+1/2)}.
\eea
By making use of the identity
\bea
\Dk (\alpha_1)\,\Dk (\alpha_2)=e^{(\alpha_1 \alpha^*_2-\alpha^*_1 \alpha_2)/2}\Dk (\alpha_1+\alpha_2),
\eea
we will find
\bea
\la m|\hat{V}|n\ra &=& e^{-i\zk (t) (m+1/2)}e^{\half (\xik\alfak^*e^{-i\zk}-\xik^* \alfak e^{i\zk})}\la m|\Dk (\xik+\alfak e^{i\zk})|n\ra,\nn\\
\eea
therefore,
\bea
\Gnkn &=& e^{-i\nu\hbar\,\ok(t)|\alfak|^2}\,e^{-i\nu\hbar\,\ok (0) (n+1/2)}\nn\\
      && \times \la n|\Dkd (\underbrace{\xik+\alfak e^{i\zk}}_{\etak})e^{i\nu H^{\kb}_0 (t)}\Dk (\underbrace{\xik+\alfak e^{i\zk}}_{\etak})|n\ra,
\eea
finally, using the formulas
\bea
e^{-\etak^* \ak} e^{i\nu \hat{H}^{\kb}_0 (t)} &=& e^{i\nu \hat{H}^{\kb}_0 (t)} e^{-\etak^* e^{i\nu\hbar\ok (t)}\ak},\nn\\
e^{\etak \akd} e^{i\nu \hat{H}^{\kb}_0 (t)} &=& e^{i\nu \hat{H}^{\kb}_0 (t)} e^{\etak e^{-i\nu\hbar\ok (t)}\akd},\nn\\
e^{\etak \akd-\etak^* \ak} &=& e^{\etak \akd}e^{-\etak^* \ak} e^{-|\etak|^2/2},\nn\\
\la n|e^{\etak \akd} &=& \sum\limits_{p=0}^n \frac{(\etak)^{n-p}}{(n-p)!}\sqrt{\frac{n!}{p!}}\,\la p|,
\eea
and \cite{Math}
\bea
\sum\limits_{p=0}^n \frac{n! (-|\etak|^2)^{n-p}}{p! [(n-p)!]^2}=L_n (|\etak|^2),
\eea
one finds
\bea
\Gnkn &=& e^{-i\nu\hbar\,\ok(t)|\alfak|^2}\,e^{i\nu\hbar\,(\ok (t)-\ok (0)) (n+1/2)}\nn\\
     && \times \,e^{\etak|^2 (e^{i\nu\hbar\ok (t)}-1)}\,L_n \big(4|\etak|^2\sin^2(\frac{\nu\hbar\ok (t)}{2})\big),
\eea
where $L_n (x)$ is the Laguerre polynomial of degree $n$.
\section{Proof of Eqs. (\ref{PkW}) and (\ref{28})}\label{ApD}
\subsection{Eq.(\ref{PkW}}
From Eq. (\ref{wdf} we have
\bea
\Gnkn &=& e^{-i\nu\hbar\ok (t)|\alfak|^2}\,e^{-|\etak|^2}\,e^{i\nu\hbar\triangle_{\kb} (t)(n+1/2)}\sum_{r=0}^\infty \frac{|\etak|^{2r}
\,e^{i r \nu\hbar\ok(t)}}{r!}\nn\\
&& \times\,\sum_{l=0}^{n}\binom{n}{l}\frac{|\etak|^{2l}}{l!}\,(e^{i\nu\hbar\ok(t)/2}-e^{-i\nu\hbar\ok(t)/2})^{2l},\nn\\
&=& e^{-|\etak|^2}e^{-i\nu\hbar\ok (t)|\alfak|^2}e^{i\nu\hbar\triangle_{\kb} (t)(n+1/2)}\sum_{r=0}^\infty \frac{|\etak|^{2r}
\,e^{i r \nu\hbar\ok(t)}}{r!}\nn\\
&& \times\,\sum_{l=0}^{n}\binom{n}{l}\frac{|\etak|^{2l}}{l!}\,e^{il\nu\hbar\ok(t)}(1-e^{-i\nu\hbar\ok(t)})^{2l},\nn\\
&=& \sum_{r=0}^\infty \sum_{l=0}^{n}\sum_{p=0}^{2l} e^{-|\etak|^2}\frac{|\etak|^{2(r+l)}}{r!\,l!}\binom{n}{l}\binom{2l}{p}(-1)^p \nn\\
&& \times\, e^{i\nu\Big((r+l-p)\hbar\ok(t)-\hbar\ok(t)|\alfak|^2+\hbar\triangle_{\kb} (t)(n+1/2)\Big)}.
\eea
Now by taking the inverse Fourier transform (see Eq. (\ref{20})) and a redefinition of variables $s=r+l-p$, we recover Eq. (\ref{PkW}).
\subsection{Eq.(\ref{28})}
For a thermal state we have
\bea
\rok (0) &=& \rokt (0)=\sum_n p_n |n\ra \la n|,
\eea
where
\bea
p_n &=& \frac{e^{-\beta\hbar\ok (0) (n+\half)}}{\sum\limits_{n=0}^\infty e^{-\beta\hbar\ok (0) (n+\half)}},\nn\\
    &=& (1-e^{-\beta\hbar\ok (0)})\,e^{-n\beta\hbar\ok (0)},
\eea
therefore,
\bea
\Gnkth &=& e^{-i\nu\hbar\,\ok(t)|\alfak|^2}\,e^{|\etak|^2 (e^{i\nu\hbar\ok (t)}-1)}\,e^{i\frac{\nu\hbar\,\triangle_{\kb} (t)}{2}}(1-e^{-\beta\hbar\ok (0)})\nn\\
     && \times\,\sum_{n=0}^\infty e^{-n(\beta\hbar\ok (0)-i\nu\hbar\triangle_{\kb} (t))}\,L_n \big(4|\etak|^2\sin^2(\frac{\nu\hbar\ok (t)}{2})\big).
\eea
Now using the identity \cite{Math}
\bea
\sum\limits_{n=0}^\infty e^{-ny}\,L_n (x)= \frac{e^{\frac{x}{1-e^{y}}}}{1-e^{-y}},
\eea
we finally find Eq. (\ref{28}).
\section{Proof of Eq. (\ref{31})}\label{ApE}
For notational simplicity, and with no lose of generality, let us replace the index $\kb$ of mutually-independent Hamiltonians ($\hat{H}^{\kb}$) by a single index $j$ as $\hat{H}_j$. Let $|E_{k_j}\ra$ be an eigenket of the Hamiltonian $\hat{H}_j (0)$ with eigenvalue $E_{k_j}$ and $|E'_{k_j}\ra$ be an eigenket of the Hamiltonian $\hat{H}_j (t)$ with eigenvalue $E'_{k_j}$. Also, let the probability that the jth subsystem being initially at the eigenket $|E_{i_j}\ra$ be denoted by $P_j (E_{i_j})$. Then, from the definition of work distribution function we can write
\bea\label{C1}
P(W) &=& \sum_{i_1,k_1,\cdots, i_j,k_j,\cdots} P_1 (E_{i_1}) |\la E_{k_1}|U_1(t)|E_{i_1}\ra|^2\,\times\cdots\nn\\
&\times & P_j (E_{i_j})|\la E_{k_n}|U_1(t)|E_{i_n}\ra|^2 \times\cdots\nn\\
&\times & \delta \bigg(W-\sum\limits_j [E_{k_j}-E_{i_j}]\bigg).
\eea
The corresponding characteristic function $G(\nu,t)$ is the Fourier transform of $P(W)$
\bea\label{C2}
G(\nu,t) = \int dW\,e^{i\nu W} P(W).
\eea
By inserting Eq. (\ref{C1}) into Eq. (\ref{C2}) and doing the integral over $W$, we find
\bea
&& G(\nu,t) = \underbrace{\sum_{i_1,k_1} P_1 (E_{i_1})|\la E_{k_1}|\hat{U}_1(t)|E_{i_1}\ra|^2\,e^{i\nu(E_{k_1}-E_{i_1})}}_{G_1 (\nu,t)}\times\cdots\nn\\
&\times & \underbrace{\sum_{i_j,k_j} P_j (E_{i_j})|\la E_{k_j}|\hat{U}_j(t)|E_{i_j}\ra|^2\,e^{i\nu(E_{k_j}-E_{i_j})}}_{G_j (\nu,t)}\times\cdots,
\eea
therefore,
\bea
G(\nu,t)=\prod\limits_j G_j(\nu,t).
\eea
\section*{References}


\begin{thebibliography}{00}
\bibitem{Jar2} C. Jarzynski, Phys. Rev. Lett. 78, 2690 (1997).
\bibitem{Jar4} C. Jarzynski, C. R. Phys. 8, 495 (2007).
\bibitem{Bochkov} G. N. Bochkov and Yu. E. Kuzovlev, Sov. Phys. JETP 45, 125 (1977).
\bibitem{Evans} D. J. Evans, E. G. D. Cohen, and G. P. Morriss, Phys. Rev. Lett. 71, 2401 (1993).
\bibitem{Jar1} C. Jarzynski, Annu. Rev. Condens. Matter Phys. 2, 329 (2011).
\bibitem{Jar3} C. Jarzynski, J. Stat. Mech. (2004) P09005.
\bibitem{Crooks} G. E. Crooks, Phys. Rev. E 60, 2721 (1999).
\bibitem{Collin} D. Collin, F. Ritort, C. Jarzynski, S. B. Smith, I. Tinoco,
and C. Bustamante, Nature (London) 437, 231 (2005);
A. N. Gupta, A. Vincent, K. Neupane, H. Yu, F. Wang, and
M. T. Woodside, Nat. Phys. 7, 631 (2011).
\bibitem{Ex1} F. Douarche, S. Ciliberto, A. Petrosyan, and I. Rabbiosi, Europhys. Lett. 70, 593 (2005).
\bibitem{Ex2} C. Bustamante, J. Liphardt, and F. Ritort, Phys. Today 58 7, 43 (2005).
\bibitem{Ex3} V. Blickle, T. Speck, L. Helden, U. Seifert, and C. Bechinger, Phys. Rev. Lett. 96, 070603 (2006).
\bibitem{Q1} H. Tasaki, e-print arXiv:cond-mat/0009244.
\bibitem{Q2} S. Mukamel, Phys. Rev. Lett. 90, 170604 (2003).
\bibitem{Q3} W. De Roeck and C. Maes, Phys. Rev. E 69, 026115 (2004).
\bibitem{Q4} M. Esposito and S. Mukamel, Phys. Rev. E 73, 046129 (2006).
\bibitem{Q5} P. Talkner and P. Hänggi, J. Phys. A 40, F569 (2007).
\bibitem{Q6} P. Talkner, P. Hänggi, and M. Morillo, Phys. Rev. E 77, 051131 (2008).
\bibitem{Q7} S. Deffner and E. Lutz, Phys. Rev. E 77, 021128 (2008).
\bibitem{Campisi} M. Campisi, P. H\"{a}nggi, and P. Talkner, Rev. Mod. Phys.
83, 771 (2011).
\bibitem{Talkner} P. Talkner, E. Lutz, and P. H\"{a}nggi, Phys. Rev. E 75, 050102 (R) (2007).
\bibitem{Esposito} M. Esposito, U. Harbola, and S. Mukamel, Rev. Mod. Phys.
81, 1665 (2009).
\bibitem{Redhead} M. Redhead, Found. Phys. 25, 123 (1995).
\bibitem{Sorkin} R. D. Sorkin, arXiv:gr-qc/9302018.
\bibitem{Dowker} F. Dowker, arXiv:1111.2308.
\bibitem{Ben} D. M. T. Benincasa, L. Borsten, M. Buck, and F. Dowker, Classical Quantum Gravity 31, 075007 (2014).
\bibitem{Souza} T. B. Batalhão, A. M. Souza, L. Mazzola, R. Auccaise, R. S.
Sarthour, I. S. Oliveira, J. Goold, G. De Chiara, M. Paternostro,
and R. M. Serra, Phys. Rev. Lett. 113, 140601 (2014).
\bibitem{Ortega} A. Ortega, E. McKay, \'{A}. M. Alhambra and E. Mart\'{i}n-Mart\'{i}nez, Phys. Rev. Lett. \textbf{122}, 240604 (2019).
\bibitem{Dorner} R. Dorner, S. R. Clark, L. Heaney, R. Fazio, J. Goold, and V.
Vedral, Phys. Rev. Lett. 110, 230601 (2013).
\bibitem{Mazzola} L. Mazzola, G. D. Chiara, and M. Paternostro, Int. J.
Quantum. Inform. 12, 1461007 (2014).
\bibitem{Os1} K. Husimi, Prog. Theor. Phys. 9, 381 (1953).
\bibitem{Os2} U. Seifert, J. Phys. A 37, L517 (2004).
\bibitem{Talkner2008} Peter Talkner, P. Sekhar Burada, and Peter H\"{a}nggi, Phys. Rev. E 78, 011115 (2008).
\bibitem{Bordag} M. Bordag, G. L. Klimchitskaya, U. Mohideen, and V. M. Mostepanenko, Advances in the
Casimir Effect (Oxford University Press, Oxford, 2008).
\bibitem{Gradstein} L. S. Gradshteyn and I. M. Ryzhik, Table of Integrals, Series
and Products (Academic, San Diego, 2000).
\bibitem{Math} Wolfram Research, Inc., Mathematica, Version 12.0, Champaign, IL (2019).
\end{thebibliography}
\end{document}